\documentclass[11pt,a4paper]{article}

\usepackage{amsmath,amssymb,amsthm}
\usepackage{bbm,color,dcolumn,enumerate}
\usepackage{fancybox,fancyhdr,graphicx}
\usepackage{multirow,psfrag,pgfplots,relsize,subcaption,times}
\usepackage{tikz,tikz-3dplot}
\usepackage[affil-it]{authblk}
\usepackage[sort,numbers]{natbib}
\usepackage[left=2cm,right=2cm,top=2cm,bottom=2cm]{geometry}

\tikzstyle{doublearr} = [latex-latex,red, line width=0.5pt]
\tikzstyle{doublearr2} = [latex-latex,green!80!black, line width=0.5pt]
\tikzstyle{mybox} = [draw=black, thick, minimum height=.6cm, rectangle, text centered]
\tikzstyle{fancytitle} = [fill=blue, text=white]

\usepackage[linktocpage=true]{hyperref}
\hypersetup{
 colorlinks,
 linkcolor=black,          
 citecolor=black,       
 filecolor=black,   
 urlcolor=black, 
 pdftitle={},
 pdfauthor={},
 pdfcreator={},
 pdfsubject={},
 pdfkeywords={}
}

\definecolor{mygray}{rgb}{0.7,0.7,0.7}

\def\R{{\mathbb{R}}}

\newcommand{\bQ}{\bar{Q}}

\newcommand{\hQ}{\hat{Q}}

\newcommand{\N}{\mathbbm{N}}

\newcommand{\GG}{\mathbf{G}}
\newcommand{\qq}{\mathbf{q}}
\newcommand{\QQ}{\mathbf{Q}}

\newcommand{\cS}{\mathcal{S}}

\newcommand{\ER}{\mathrm{ER}}

\newcommand{\dis}{\text{\textsc{dis}}}
\newcommand{\com}{\text{\textsc{com}}}

\newcommand{\CJSQ}{\mbox{CJSQ}}

\newcommand{\dif}{\ensuremath{\mbox{d}}}  

\newcommand\ind[1]{\ensuremath{\mathbbm{1}_{\left[#1\right]}}} 

\providecommand{\href}[2]{#2}

\newtheorem{theorem}{Theorem}[section]

\newtheorem{corollary}[theorem]{Corollary}
\newtheorem{definition}[theorem]{Definition}

\theoremstyle{definition}

\makeatletter
\renewcommand{\fnum@figure}{\small\textbf{\figurename~\thefigure}}
\makeatother

\makeatletter
\renewcommand{\fnum@table}{\small\textbf{\tablename~\thetable}}
\makeatother

\numberwithin{equation}{section}
\numberwithin{theorem}{section}

\graphicspath{{./},{./images/}}

\usepackage{array}   
\newcolumntype{L}[1]{>{\raggedright\let\newline\\\arraybackslash\hspace{0pt}}m{#1}}
\newcolumntype{C}[1]{>{\centering\let\newline\\\arraybackslash\hspace{0pt}}m{#1}}
\newcolumntype{R}[1]{>{\raggedleft\let\newline\\\arraybackslash\hspace{0pt}}m{#1}}

\begin{document}

\title{Scalable Load Balancing in Networked Systems: \\
Universality Properties and Stochastic Coupling Methods}
\author[1]{Mark van der Boor}
\author[1,2]{Sem C.~Borst}
\author[1]{Johan S.H.~van Leeuwaarden}
\author[1]{Debankur Mukherjee}
\affil[1]{Eindhoven University of Technology, The Netherlands}
\affil[2]{Nokia Bell Labs, Murray Hill, NJ, USA}

\date{\today}

\maketitle

\begin{abstract}
We present an overview of scalable load balancing algorithms
which provide favorable delay performance in large-scale systems,
and yet only require minimal implementation overhead.
Aimed at a broad audience, the paper starts with an introduction
to the basic load balancing scenario -- referred to as the
\emph{supermarket model} -- consisting of a single dispatcher where
tasks arrive that must immediately be forwarded to one
of $N$~single-server queues.
The supermarket model is a dynamic counterpart of the classical
balls-and-bins setup where balls must be sequentially distributed
across bins.

A popular class of load balancing algorithms are so-called
power-of-$d$ or JSQ($d$) policies, where an incoming task is assigned
to a server with the shortest queue among $d$~servers selected
uniformly at random.
As the name reflects, this class includes the celebrated
Join-the-Shortest-Queue (JSQ) policy as a special case ($d = N$), which
has strong stochastic optimality properties and yields a mean
waiting time that \emph{vanishes} as $N$ grows large for any fixed
subcritical load.
However, a nominal implementation of the JSQ policy involves
a prohibitive communication burden in large-scale deployments.
In contrast, a simple random assignment policy ($d = 1$) does not
entail any communication overhead, but the mean waiting time remains
constant as $N$ grows large for any fixed positive load.

In order to examine the fundamental trade-off between delay
performance and implementation overhead, we consider an asymptotic
regime where the diversity parameter $d(N)$ depends on~$N$.
We investigate what growth rate of $d(N)$ is required to match the
optimal performance of the JSQ policy on fluid and diffusion scale,
and achieve a vanishing waiting time in the limit.
The results demonstrate that the asymptotics for the JSQ($d(N)$)
policy are insensitive to the exact growth rate of $d(N)$,
as long as the latter is sufficiently fast, implying that the
optimality of the JSQ policy can asymptotically be preserved
while dramatically reducing the communication overhead.

Stochastic coupling techniques play an instrumental role
in establishing the asymptotic optimality and universality properties,
and augmentations of the coupling constructions allow these properties
to be extended to infinite-server settings and network scenarios.
We additionally show how the communication overhead can be reduced
yet further by the so-called Join-the-Idle-Queue (JIQ) scheme,
leveraging memory at the dispatcher to keep track of idle servers.
\end{abstract}

\newpage

\section{Introduction}

In the present paper we review scalable load balancing algorithms
(LBAs) which achieve excellent delay performance in large-scale
systems and yet only involve low implementation overhead.
LBAs play a critical role in distributing service requests or tasks
(e.g.~compute jobs, data base look-ups, file transfers) among servers
or distributed resources in parallel-processing systems.
The analysis and design of LBAs has attracted strong attention in recent
years, mainly spurred by crucial scalability challenges arising
in cloud networks and data centers with massive numbers of servers.

LBAs can be broadly categorized as static, dynamic, or some intermediate
blend, depending on the amount of feedback or state information
(e.g.~congestion levels) that is used in allocating tasks.
The use of state information naturally allows dynamic policies
to achieve better delay performance, but also involves higher
implementation complexity and a substantial communication burden.
The latter issue is particularly pertinent in cloud networks and data
centers with immense numbers of servers handling a huge influx
of service requests.
In order to capture the large-scale context, we examine scalability
properties through the prism of asymptotic scalings where the system
size grows large, and identify LBAs which strike an optimal balance
between delay performance and implementation overhead in that regime.

The most basic load balancing scenario consists of $N$~identical
parallel servers and a dispatcher where tasks arrive that must
immediately be forwarded to one of the servers.
Tasks are assumed to have unit-mean exponentially distributed service
requirements, and the service discipline at each server is supposed
to be oblivious to the actual service requirements.
In this canonical setup, the celebrated Join-the-Shortest-Queue (JSQ)
policy has several strong stochastic optimality properties.
In particular, the JSQ policy achieves the minimum mean overall
delay among all non-anticipating policies that do not have any
advance knowledge of the service requirements \cite{EVW80,Winston77}.
In order to implement the JSQ policy however, a dispatcher requires
instantaneous knowledge of all the queue lengths, which may involve
a prohibitive communication burden with a large number of servers~$N$.

This poor scalability has motivated consideration of JSQ($d$) policies,
where an incoming task is assigned to a server with the shortest queue
among $d \geq 2$ servers selected uniformly at random.
Note that this involves exchange of $2 d$ messages per task,
irrespective of the number of servers~$N$.
Results in Mitzenmacher~\cite{Mitzenmacher01} and Vvedenskaya
{\em et al.}~\cite{VDK96} indicate that even sampling as few as $d = 2$
servers yields significant performance enhancements over purely
random assignment ($d = 1$) as $N$ grows large, which is commonly
referred to as the ``power-of-two'' or ``power-of-choice'' effect.
Specifically, when tasks arrive at rate $\lambda N$,
the queue length distribution at each individual server exhibits
super-exponential decay for any fixed $\lambda < 1$ as $N$ grows large,
compared to exponential decay for purely random assignment.

As illustrated by the above, the diversity parameter~$d$ induces
a fundamental trade-off between the amount of communication overhead
and the delay performance.
Specifically, a random assignment policy does not entail any
communication burden, but the mean waiting time remains \emph{constant}
as $N$ grows large for any fixed $\lambda > 0$.
In contrast, a nominal implementation of the JSQ policy (without
maintaining state information at the dispatcher) involves $2 N$
messages per task, but the mean waiting time \emph{vanishes}
as $N$ grows large for any fixed $\lambda < 1$.
Although JSQ($d$) policies with $d \geq 2$ yield major performance
improvements over purely random assignment while reducing the
communication burden by a factor O($N$) compared to the JSQ policy,
the mean waiting time \emph{does not vanish} in the limit.
Thus, no fixed value of~$d$ will provide asymptotically optimal
delay performance.
This is evidenced by results of Gamarnik {\em et al.}~\cite{GTZ16}
indicating that in the absence of any memory at the dispatcher the
communication overhead per task \emph{must increase} with~$N$ in order
for any scheme to achieve a zero mean waiting time in the limit.

We will explore the intrinsic trade-off between delay performance
and communication overhead as governed by the diversity parameter~$d$,
in conjunction with the relative load~$\lambda$.
The latter trade-off is examined in an asymptotic regime where not
only the overall task arrival rate is assumed to grow with~$N$,
but also the diversity parameter is allowed to depend on~$N$.
We write $\lambda(N)$ and $d(N)$, respectively, to explicitly reflect
that, and investigate what growth rate of $d(N)$ is required,
depending on the scaling behavior of $\lambda(N)$, in order to achieve
a zero mean waiting time in the limit.
We establish that the fluid-scale and diffusion-scale limiting
processes are insensitive to the exact growth rate of $d(N)$,
as long as the latter is sufficiently fast, and in particular coincide
with the limiting processes for the JSQ policy.
This reflects a remarkable universality property and demonstrates
that the optimality of the JSQ policy can asymptotically be preserved 
while dramatically lowering the communication overhead.

We will extend the above-mentioned universality properties to network
scenarios where the $N$~servers are assumed to be inter-connected
by some underlying graph topology $G_N$.
Tasks arrive at the various servers as independent Poisson processes
of rate~$\lambda$, and each incoming task is assigned to whichever
server has the shortest queue among the one where it appears and its
neighbors in $G_N$.
In case $G_N$ is a clique, each incoming task is assigned to the
server with the shortest queue across the entire system,
and the behavior is equivalent to that under the JSQ policy.
The above-mentioned stochastic optimality properties of the JSQ policy
thus imply that the queue length process in a clique will be `better'
than in an arbitrary graph~$G_N$.
We will establish sufficient conditions for the fluid-scaled
and diffucion-scaled versions of the queue length process in
an arbitrary graph to be equivalent to the limiting processes
in a clique as $N \to \infty$.
The conditions reflect similar universality properties as described above,
and in particular demonstrate that the optimality of a clique can
asymptotically be preserved while markedly reducing the number
of connections, provided the graph $G_N$ is suitably random.

While a zero waiting time can be achieved in the limit by sampling
only $d(N) = o(N)$ servers, the amount of communication overhead
in terms of $d(N)$ must still grow with~$N$.
This may be explained from the fact that a large number of servers
need to be sampled for each incoming task to ensure that at least one
of them is found idle with high probability.
As alluded to above, this can be avoided by introducing memory
at the dispatcher, in particular maintaining a record of vacant servers,
and assigning tasks to idle servers, if there are any.
This so-called Join-the-Idle-Queue (JIQ) scheme \cite{BB08,LXKGLG11}
has gained huge popularity recently, and can be implemented through
a simple token-based mechanism generating at most one message per task.
As established by Stolyar~\cite{Stolyar15}, the fluid-scaled queue
length process under the JIQ scheme is equivalent to that under the
JSQ policy as $N \to \infty$, and this result can be shown to extend
the diffusion-scaled queue length process.
Thus, the use of memory allows the JIQ scheme to achieve asymptotically
optimal delay performance with minimal communication overhead.
In particular, ensuring that tasks are assigned to idle servers
whenever available is sufficient to achieve asymptotic optimality,
and using any additional queue length information yields no meaningful
performance benefits on the fluid or diffusion levels.

Stochastic coupling techniques play an instrumental role in the proofs
of the above-described universality and asymptotic optimality properties.
A direct analysis of the queue length processes under a JSQ($d(N)$)
policy, in a load balancing graph $G_N$, or under the JIQ scheme
is confronted with unsurmountable obstacles.
As an alternative route, we leverage novel stochastic coupling
constructions to relate the relevant queue length processes to the
corresponding processes under a JSQ policy, and show that the
deviation between these two is asymptotically negligible under mild
assumptions on $d(N)$ or $G_N$.

While the stochastic coupling schemes provide a remarkably effective
and overarching approach, they defy a systematic recipe and involve
some degree of ingenuity and customization.
Indeed, the specific coupling arguments that we develop are not only
different from those that were originally used in establishing the
stochastic optimality properties of the JSQ policy, but also differ
in critical ways between a JSQ($d(N)$) policy, a load balancing graph
$G_N$, and the JIQ scheme.
Yet different coupling constructions are devised for model variants
with infinite-server dynamics that we will discuss in Section~\ref{bloc}.

The remainder of the paper is organized as follows.
In Section~\ref{spec} we discuss a wide spectrum of LBAs and evaluate
their scalability properties.
In Section~\ref{jsqd} we introduce some useful preliminaries,
review fluid and diffusion limits for the JSQ policy as well as
JSQ($d$) policies with a fixed value of~$d$, and explore the trade-off
between delay performance and communication overhead as function
of the diversity parameter~$d$.
In particular, we establish asymptotic universality properties
for JSQ($d$) policies, which are extended to systems with server pools
and network scenarios in Sections~\ref{bloc} and~\ref{networks},
respectively.
In Section~\ref{token} we establish asymptotic optimality properties
for the JIQ scheme.
We discuss somewhat related redundancy policies and alternative scaling
regimes and performance metrics in Section~\ref{miscellaneous}.

\section{Scalability spectrum}
\label{spec}

In this section we review a wide spectrum of LBAs and examine their
scalability properties in terms of the delay performance vis-a-vis
the associated implementation overhead in large-scale systems.

\subsection{Basic model}

Throughout this section and most of the paper, we focus on a basic
scenario with $N$ parallel single-server infinite-buffer queues
and a single dispatcher where tasks arrive as a Poisson process
of rate~$\lambda(N)$, as depicted in Figure~\ref{figJSQ}.
Arriving tasks cannot be queued at the dispatcher,
and must immediately be forwarded to one of the servers.
This canonical setup is commonly dubbed the \emph{supermarket model}.
Tasks are assumed to have unit-mean exponentially distributed service
requirements, and the service discipline at each server is supposed
to be oblivious to the actual service requirements.

In Section~\ref{bloc} we consider some model variants with $N$~server
pools and possibly finite buffers and in Section~\ref{networks}
we will treat network generalizations of the above model.

\subsection{Asymptotic scaling regimes}
\label{asym}

An exact analysis of the delay performance is quite involved,
if not intractable, for all but the simplest LBAs.
Numerical evaluation or simulation are not straightforward either,
especially for high load levels and large system sizes.
A common approach is therefore to consider various limit regimes,
which not only provide mathematical tractability and illuminate the
fundamental behavior, but are also natural in view of the typical
conditions in which cloud networks and data centers operate.
One can distinguish several asymptotic scalings that have been used
for these purposes:
(i) In the classical heavy-traffic regime, $\lambda(N) = \lambda N$
with a fixed number of servers~$N$ and a relative load~$\lambda$
that tends to one in the limit.
(ii) In the conventional large-capacity or many-server regime,
the relative load $\lambda(N) / N$ approaches a constant $\lambda < 1$
as the number of servers~$N$ grows large.
(iii) The popular Halfin-Whitt regime~\cite{HW81} combines heavy traffic
with a large capacity, with
\begin{equation}
\label{eq:HW}
\frac{N - \lambda(N)}{\sqrt{N}} \to \beta > 0 \mbox{ as } N \to \infty,
\end{equation}
so the relative capacity slack behaves as $\beta / \sqrt{N}$
as the number of servers~$N$ grows large.
(iv) The so-called non-degenerate slow-down regime~\cite{Atar12}
involves $N - \lambda(N) \to \gamma > 0$, so the relative capacity
slack shrinks as $\gamma / N$ as the number of servers~$N$ grows large.

The term non-degenerate slow-down refers to the fact that in the
context of a centralized multi-server queue, the mean waiting time
in regime (iv) tends to a strictly positive constant as $N \to \infty$,
and is thus of similar magnitude as the mean service requirement.
In contrast, in regimes (ii) and (iii), the mean waiting time
decays exponentially fast in~$N$ or is of the order~$1 / \sqrt{N}$,
respectively, as $N \to \infty$, while in regime (i) the mean waiting
time grows arbitrarily large relative to the mean service requirement.

In the present paper we will focus on scalings (ii) and (iii),
and occasionally also refer to these as fluid and diffusion scalings, 
since it is natural to analyze the relevant queue length process
on fluid scale ($1 / N$) and diffusion scale ($1 / \sqrt{N}$)
in these regimes, respectively.
We will not provide a detailed account of scalings (i) and (iv),
which do not capture the large-scale perspective and do not allow
for low delays, respectively, but we will briefly revisit these
regimes in Section~\ref{miscellaneous}.

\subsection{Random assignment: N independent M/M/1 queues}
\label{random}

One of the most basic LBAs is to assign each arriving task to a server
selected uniformly at random.
In that case, the various queues collectively behave as
$N$~independent M/M/1 queues, each with arrival rate $\lambda(N) / N$
and unit service rate.
In particular, at each of the queues, the total number of tasks in
stationarity has a geometric distribution with parameter $\lambda(N) / N$.
By virtue of the PASTA property, the probability that an arriving task
incurs a non-zero waiting time is $\lambda(N) / N$.
The mean number of waiting tasks (excluding the possible task in service)
at each of the queues is $\frac{\lambda(N)^2}{N (N - \lambda(N))}$,
so the total mean number of waiting tasks is
$\frac{\lambda(N)^2}{N - \lambda(N)}$, which by Little's law implies that
the mean waiting time of a task is $\frac{\lambda(N)}{N - \lambda(N)}$.
In particular, when $\lambda(N) = N \lambda$, the probability that
a task incurs a non-zero waiting time is $\lambda$,
and the mean waiting time of a task is $\frac{\lambda}{1 - \lambda}$,
independent of~$N$, reflecting the independence of the various queues.

A slightly better LBA is to assign tasks to the servers in
a Round-Robin manner, dispatching every $N$-th task to the same server.
In the large-capacity regime where $\lambda(N) = N \lambda$,
the inter-arrival time of tasks at each given queue will then converge
to a constant $1 / \lambda$ as $N \to \infty$.
Thus each of the queues will behave as an D/M/1 queue in the limit,
and the probability of a non-zero waiting time and the mean waiting
time will be somewhat lower than under purely random assignment.
However, both the probability of a non-zero waiting time and the mean
waiting time will still tend to strictly positive values and not vanish
as $N \to \infty$.

\subsection{Join-the-Shortest Queue (JSQ)}
\label{ssec:jsq}

Under the Join-the-Shortest-Queue (JSQ) policy, each arriving task is
assigned to the server with the currently shortest queue (ties are
broken arbitrarily).
In the basic model described above, the JSQ policy has several strong
stochastic optimality properties, and yields the `most balanced
and smallest' queue process among all non-anticipating policies that
do not have any advance knowledge of the service requirements
\cite{EVW80,Winston77}.
Specifically, the JSQ policy minimizes the joint queue length vector
in a stochastic majorization sense, and in particular stochastically
minimizes the total number of tasks in the system, and hence the mean
overall delay.
In order to implement the JSQ policy however, a dispatcher requires
instantaneous knowledge of the queue lengths at all the servers.
A nominal implementation would involve exchange of $2 N$ messages per task,
and thus yield a prohibitive communication burden in large-scale systems.

\subsection{Join-the-Smallest-Workload (JSW): centralized M/M/N queue}
\label{ssec:jsw}

Under the Join-the-Smallest-Workload (JSW) policy, each arriving task
is assigned to the server with the currently smallest workload.
Note that this is an anticipating policy, since it requires advance
knowledge of the service requirements of all the tasks in the system.
Further observe that this policy (myopically) minimizes the waiting
time for each incoming task, and mimics the operation of a centralized
$N$-server queue with a FCFS discipline.
The equivalence with a centralized $N$-server queue yields a strong
optimality property of the JSW policy:
The vector of joint workloads at the various servers observed by each
incoming task is smaller in the Schur convex sense than under any
alternative admissible policy~\cite{FC01}.

The equivalence with a centralized FCFS queue means that there cannot
be any idle servers while tasks are waiting.
In our setting with Poisson arrivals and exponential service
requirements, it can therefore be shown that the total number of tasks
under the JSW policy is stochastically smaller than under the JSQ policy.
At the same time, it means that the total number of tasks under the
JSW policy behaves as a birth-death process, which renders it far more
tractable than the JSQ policy.
Specifically, given that all the servers are busy, the total number
of waiting tasks is geometrically distributed with parameter
$\lambda(N) / N$.
Thus the total mean number of waiting tasks is
$\Pi_W(N, \lambda(N)) \frac{\lambda(N)}{N - \lambda(N)}$,
and the mean waiting time is
$\Pi_W(N, \lambda(N)) \frac{1}{N - \lambda(N)}$,
with $\Pi_W(N, \lambda(N)$ denoting the probability of all servers
being occupied and a task incurring a non-zero waiting time.
This immediately shows that the mean waiting time is smaller
by at least a factor $\lambda(N)$ than for the random assignment
policy considered in Subsection~\ref{random}.

In the large-capacity regime $\lambda(N) = N \lambda$, it can be shown
that the probability $\Pi_W(N, \lambda(N))$ of a non-zero waiting time
decays exponentially fast in~$N$, and hence so does the mean waiting time.
In the Halfin-Whitt heavy-traffic regime~\eqref{eq:HW},
the probability $\Pi_W(N, \lambda(N))$ of a non-zero waiting time
converges to a finite constant $\Pi_W^\star(\beta)$,
implying that the mean waiting time of a task is of the order
$1 / \sqrt{N}$, and thus vanishes as $N \to \infty$.

\subsection{Power-of-d load balancing (JSQ(d))}
\label{ssec:powerd}

As mentioned above, the achilles heel of the JSQ policy is its
excessive communication overhead in large-scale systems.
This poor scalability has motivated consideration of so-called
JSQ($d$) policies, where an incoming task is assigned to a server
with the shortest queue among $d$~servers selected uniformly at random.
Results in Mitzenmacher~\cite{Mitzenmacher01} and Vvedenskaya
{\em et al.}~\cite{VDK96} indicate that even sampling as few as $d = 2$
servers yields significant performance enhancements over purely
random assignment ($d = 1$) as $N \to \infty$.
Specifically, in the fluid regime where $\lambda(N) = \lambda N$,
the probability that there are $i$ or more tasks at a given queue is
proportional to $\lambda^{\frac{d^i - 1}{d - 1}}$ as $N \to \infty$,
and thus exhibits super-exponential decay as opposed to exponential decay
for the random assignment policy considered in Subsection~\ref{random}.

As illustrated by the above, the diversity parameter~$d$ induces
a fundamental trade-off between the amount of communication overhead
and the performance in terms of queue lengths and delays.
A rudimentary implementation of the JSQ policy ($d = N$, without
replacement) involves $O(N)$ communication overhead per task,
but it can be shown that the probability of a non-zero waiting time
and the mean waiting \emph{vanish} as $N \to \infty$, just like
in a centralized queue.
Although JSQ($d$) policies with a fixed parameter $d \geq 2$ yield
major performance improvements over purely random assignment
while reducing the communication burden by a factor O($N$) compared
to the JSQ policy, the probability of a non-zero waiting time
and the mean waiting time \emph{do not vanish} as $N \to \infty$.

In Subsection~\ref{univ} we will explore the intrinsic trade-off
between delay performance and communication overhead as function
of the diversity parameter~$d$, in conjunction with the relative load.
We will examine an asymptotic regime where not only the total task
arrival rate $\lambda(N)$ is assumed to grow with~$N$,
but also the diversity parameter is allowed to depend on~$N$.
As will be demonstrated, the optimality of the JSQ policy ($d(N) = N$)
can be preserved, and in particular a vanishing waiting time can be
achieved in the limit as $N \to \infty$, even when $d(N) = o(N)$,
thus dramatically lowering the communication overhead. 

\subsection{Token-based strategies: Join-the-Idle-Queue (JIQ)}
\label{ssec:jiq}

While a zero waiting time can be achieved in the limit by sampling
only $d(N) = o(N)$ servers, the amount of communication overhead
in terms of $d(N)$ must still grow with~$N$.
This can be countered by introducing memory at the dispatcher,
in particular maintaining a record of vacant servers,
and assigning tasks to idle servers as long as there are any,
or to a uniformly at random selected server otherwise.
This so-called Join-the-Idle-Queue (JIQ) scheme \cite{BB08,LXKGLG11}
has received keen interest recently, and can be implemented through
a simple token-based mechanism.
Specifically, idle servers send tokens to the dispatcher to advertise
their availability, and when a task arrives and the dispatcher has
tokens available, it assigns the task to one of the corresponding
servers (and disposes of the token).
Note that a server only issues a token when a task completion leaves
its queue empty, thus generating at most one message per task.
Surprisingly, the mean waiting time and the probability of a non-zero
waiting time vanish under the JIQ scheme in both the fluid and diffusion
regimes, as we will further discuss in Section~\ref{token}.
Thus, the use of memory allows the JIQ scheme to achieve asymptotically
optimal delay performance with minimal communication overhead.

\subsection{Performance comparison}
\label{ssec:perfcomp}

\begin{figure}
\includegraphics[width=\linewidth]{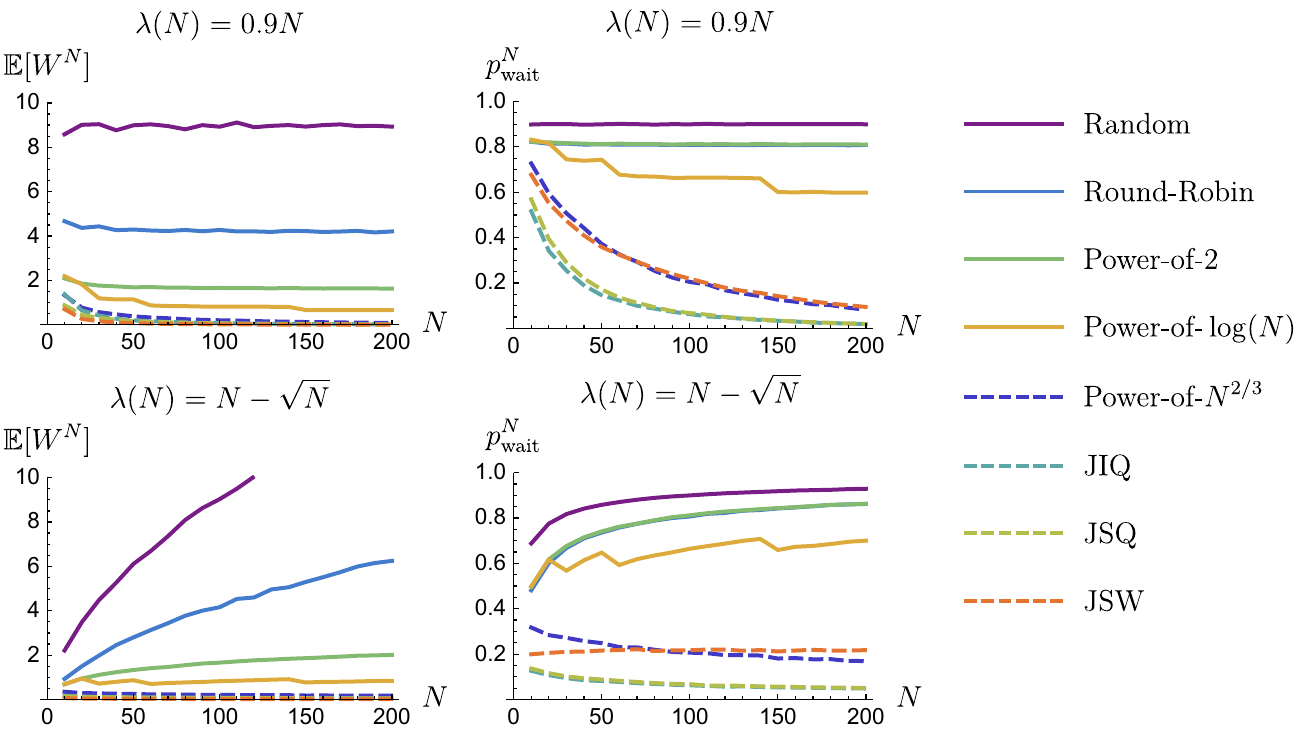}
\caption{Simulation results for mean waiting time $\mathbb{E}[W^N]$
and probability of a non-zero waiting time $p_{\textup{wait}}^N$,
for both a fluid regime and a diffusion regime.}
\label{differentschemes}
\end{figure}

We now present some simulation experiments that we have conducted
to compare the above-described LBAs in terms of delay performance.
Specifically, we evaluate the mean waiting time and the probability
of a non-zero waiting time in both a fluid regime ($\lambda(N) = 0.9 N$)
and a diffusion regime ($\lambda(N) = N - \sqrt{N}$).
The results are shown in Figure~\ref{differentschemes}.
We are especially interested in distinguishing two classes of LBAs --
ones delivering a mean waiting time and probability of a non-zero
waiting time that vanish asymptotically, and ones that fail to do so
-- and relating that dichotomy to the associated overhead.

\paragraph{JSQ, JIQ, and JSW.}

JSQ, JIQ and JSW evidently have a vanishing waiting time in both the
fluid and the diffusion regime as discussed in Subsections~\ref{ssec:jsq},
\ref{ssec:jsw} and~\ref{ssec:jiq}.
The optimality of JSW as mentioned in Subsection~\ref{ssec:jsw}
can also be clearly observed.

However, there is a significant difference between JSW and JSQ/JIQ
in the diffusion regime.
We observe that the probability of a non-zero waiting time
\emph{approaches a positive constant} for JSW,
while it \emph{vanishes} for JSQ/JIQ.
In other words, the mean of all positive waiting times is of a larger
order of magnitude in JSQ/JIQ compared to JSW.
Intuitively, this is clear since in JSQ/JIQ, when a task is placed
in a queue, it waits for at least a residual service time.
In JSW, which is equivalent to the M/M/$N$ queue, a task that cannot
start service immediately, joins a queue that is collectively drained
by all the $N$~servers

\paragraph{Random and Round-Robin.}

The mean waiting time does not vanish for Random and Round-Robin
in the fluid regime, as already mentioned in Subsection~\ref{random}.
Moreover, the mean waiting time grows without bound in the diffusion
regime for these two schemes.
This is because the system can still be decomposed, and the loads
of the individual M/M/1 and D/M/1 queues tend to~1.

\paragraph{JSQ(${\bf d}$) policies.}

Three versions of JSQ($d$) are included in the figures;
$d(N)=2\not\to \infty$, $d(N)=\lfloor\log(N)\rfloor\to \infty$
and $d(N)=N^{2/3}$ for which $\frac{d(N)}{\sqrt{N}\log(N)}\to \infty$.
Note that the graph for $d(N)=\lfloor \log(N) \rfloor$ shows sudden
jumps when $d(N)$ increases by~$1$.
The variants for which $d(N)\to\infty$ have a vanishing waiting time
in the fluid regime, while $d = 2$ does not.
The latter observation is a manifestation of the results of Gamarnik
{\em et al.}~\cite{GTZ16} mentioned in the introduction, since JSQ($d$)
uses no memory and the overhead per task does not increase with~$N$.
Furthermore, it follows that JSQ($d$) policies outperform
Random and Round-Robin, while JSQ/JIQ/JSW are better in terms of mean
waiting time. \\

In order to succinctly capture the results and observed dichotomy
in Figure~\ref{differentschemes}, we provide an overview of the delay
performance of the various LBAs and the associated overhead
in Table~\ref{table}, where $q_i^\star$ denotes the stationary fraction
of servers with $i$ or more tasks.

\begin{table}\centering
\def\arraystretch{1.5}%
\begin{tabular}{|C{3cm}|C{3cm}|C{2.5cm}|C{3cm}|C{1.5cm}|}
\hline
Scheme & Queue length & Waiting time (fixed $\lambda < 1$) &
Waiting time ($1 - \lambda \sim 1 / \sqrt{N}$) & Overhead per task \\
\hline\hline
Random & $q_i^\star = \lambda^i$ &  $\frac{\lambda}{1 - \lambda}$ &
$\Theta(\sqrt{N})$ & 0 \\
\hline
JSQ($d$) & $q_i^\star = \lambda^{\frac{d^i - 1}{d - 1}}$ &
$\Theta$(1) & $\Omega(\log{N})$ & $2 d$ \\
\hline
\mbox{$d(N)$} \mbox{$\to \infty$} & same as JSQ & same as JSQ & ?? &
$2d(N)$ \\
\hline
$\frac{d(N)}{\sqrt{N} \log(N)}\to \infty$ &
same as JSQ & same as JSQ & same as JSQ & $2d(N)$ \\
\hline
JSQ & $q_1^\star = \lambda$, $q_2^\star =$ o(1) & o(1) &
$\Theta(1 / \sqrt{N})$ & $2 N$ \\
\hline\hline
JIQ & same as JSQ & same as JSQ & same as JSQ & $\leq 1$ \\
\hline
\end{tabular}
\caption{Queue length distribution, waiting times and communication
overhead for various LBAs.}
\label{table}
\end{table}

\section{JSQ(d) policies and universality properties}
\label{jsqd}

In this section we first introduce some useful preliminary concepts,
then review fluid and diffusion limits for the JSQ policy as well as
JSQ($d$) policies with a fixed value of~$d$, and finally discuss
universality properties when the diversity parameter $d(N)$ is being
scaled with~$N$.

As described in the previous section, we focus on a basic scenario
where all the servers are homogeneous, the service requirements are
exponentially distributed, and the service discipline at each server
is oblivious of the actual service requirements.
In order to obtain a Markovian state description, it therefore
suffices to only track the number of tasks, and in fact we do not need
to keep record of the number of tasks at each individual server,
but only count the number of servers with a given number of tasks.
Specifically, we represent the state of the system by a vector
$\QQ(t) := \left(Q_1(t), Q_2(t), \dots\right)$,
with $Q_i(t)$ denoting the number of servers with $i$ or more tasks
at time~$t$, including the possible task in service, $i = 1, 2 \dots$.
Note that if we represent the queues at the various servers as (vertical)
stacks, and arrange these from left to right in non-descending order,
then the value of $Q_i$ corresponds to the width of the $i$-th (horizontal)
row, as depicted in the schematic diagram in Figure~\ref{figB}.

\begin{figure}
\begin{minipage}{\textwidth}
\begin{center}
\hfill\parbox{0.48\textwidth}{\centering
\begin{tikzpicture}[scale=0.5]
\node[above] at (3.750,5) {\small$\lambda(N)$};
\draw[thick,black,->] (2.75,5)--(4.75,5);
\draw[thick,black,fill=black](5,5) circle [radius=0.1];
\draw[thick,black,->,dash pattern=on 2 off 1] (5,5)--(7,5);
\draw[thick,black,->,dash pattern=on 2 off 1] (5,5)--(7,6.5);
\draw[thick,black,->,dash pattern=on 2 off 1] (5,5)--(7,8);
\draw[thick,black,->,dash pattern=on 2 off 1] (5,5)--(7,3.5);
\draw[thick,black,->,dash pattern=on 2 off 1] (5,5)--(7,2);
\foreach \i in {1,...,4}
{
\draw[thick,black,fill=mygray](13-\i,8) circle [radius=0.4];
}
\foreach \i in {1,...,2}
{
\draw[thick,black,fill=mygray](13-\i,6.5) circle [radius=0.4];
}
\foreach \i in {1,...,3}
{
\draw[thick,black,fill=mygray](13-\i,5) circle [radius=0.4];
}
\foreach \i in {1,...,5}
{
\draw[thick,black,fill=mygray](13-\i,2) circle [radius=0.4];
}
\draw[thick,black](13.2,8) circle [radius=0.5];
\draw[thick,black](13.2,6.5) circle [radius=0.5];
\draw[thick,black](13.2,5) circle [radius=0.5];
\draw[thick,black](13.2,2) circle [radius=0.5];
\draw[thick,black,->] (13.7,8)--(14.5,8);
\draw[thick,black,->] (13.7,6.5)--(14.5,6.5);
\draw[thick,black,->] (13.7,5)--(14.5,5);
\draw[thick,black,->] (13.7,2)--(14.5,2);
\node at (13.2,8) {\small 1};
\node at (13.2,6.5) {\small 2};
\node at (13.2,5) {\small 3};
\node at (13.2,3.5) {$\vdots$};
\node at (13.2,2) {\small $N$};
\end{tikzpicture}
}
\parbox{0.48\textwidth}{\centering
\begin{tikzpicture}[scale=.60]
\foreach \x in {10, 9,...,1}
    \draw (\x,6)--(\x,0)--(\x+.7,0)--(\x+.7,6);
\foreach \x in {10, 9,...,1}
    \draw (\x+.35,-.45) node[circle,inner sep=0pt, minimum size=10pt,draw,thick] {{{\tiny $\mathsmaller{\x}$}}} ;
\foreach \y in {0, .5}
    \draw[fill=mygray,mygray] (1.15,.1+\y) rectangle (1.55,.5+\y);
\foreach \y in {0, .5, 1, 1.5}
    \draw[fill=mygray,mygray] (2.15,.1+\y) rectangle (2.55,.5+\y);
\foreach \y in {0, .5, 1, 1.5}
    \draw[fill=mygray,mygray] (3.15,.1+\y) rectangle (3.55,.5+\y);
\foreach \y in {0, .5, 1, 1.5, 2, 2.5}
    \draw[fill=mygray,mygray] (4.15,.1+\y) rectangle (4.55,.5+\y);
\foreach \y in {0, .5, 1, 1.5, 2, 2.5, 3}
    \draw[fill=mygray,mygray] (5.15,.1+\y) rectangle (5.55,.5+\y);
\foreach \y in {0, .5, 1, 1.5, 2, 2.5, 3, 3.5, 4}
    \draw[fill=mygray,mygray] (6.15,.1+\y) rectangle (6.55,.5+\y);
\foreach \y in {0, .5, 1, 1.5, 2, 2.5, 3, 3.5, 4}
    \draw[fill=mygray,mygray] (7.15,.1+\y) rectangle (7.55,.5+\y);
\foreach \y in {0, .5, 1, 1.5, 2, 2.5, 3, 3.5, 4, 4.5, 5}
    \draw[fill=mygray,mygray] (8.15,.1+\y) rectangle (8.55,.5+\y);
\foreach \y in {0, .5, 1, 1.5, 2, 2.5, 3, 3.5, 4, 4.5, 5}
    \draw[fill=mygray,mygray] (9.15,.1+\y) rectangle (9.55,.5+\y);
\foreach \y in {0, .5, 1, 1.5, 2, 2.5, 3, 3.5, 4, 4.5, 5}
    \draw[fill=mygray,mygray] (10.15,.1+\y) rectangle (10.55,.5+\y);

\draw[thick] (.9,0) rectangle (10.8,.5);
\draw[thick] (.9,.5) rectangle (10.8,1);
\draw[thick] (3.9,2.5) rectangle (10.8,3);

\draw  (12, .2) node {{\scriptsize $\leftarrow Q_1=10$}};
\draw  (12, .9) node {{\scriptsize $\leftarrow Q_2=10$}};

\draw  (11.85, 1.3) node {{\tiny $\cdot$}};
\draw  (11.85, 1.8) node {{\tiny $\cdot$}};
\draw  (11.85, 2.3) node {{\tiny $\cdot$}};
\draw  (12, 2.7) node {{\scriptsize $\leftarrow Q_i=7$}};
\draw  (11.85, 3.3) node {{\tiny $\cdot$}};
\draw  (11.85, 3.8) node {{\tiny $\cdot$}};
\draw  (11.85, 4.3) node {{\tiny $\cdot$}};
\end{tikzpicture}
}\\[1ex]
\parbox{0.48\textwidth}{\centering
\captionof{figure}{Tasks arrive at the dispatcher as a Poisson process
of rate $\lambda(N)$, and are forwarded to one of the $N$ servers
according to some specific load balancing algorithm.
\label{figJSQ}}
}
\hfill\parbox{0.48\textwidth}{\centering
\captionof{figure}{The value of $Q_i$ represents the width of the
$i$-th row, when the servers are arranged in non-descending order
of their queue lengths.
\label{figB}}
}
\end{center}
\end{minipage}
\end{figure}

In order to examine the asymptotic behavior when the number of servers~$N$
grows large, we consider a sequence of systems indexed by~$N$,
and attach a superscript~$N$ to the associated state variables.

The fluid-scaled occupancy state is denoted by
$\qq^N(t) := (q_1^N(t), q_2^N(t), \dots)$, with $q_i^N(t) = Q_i^N(t) / N$
representing the fraction of servers in the $N$-th system with $i$
or more tasks as time~$t$, $i = 1, 2, \dots$.
Let
$\cS = \{\qq \in [0, 1]^\infty: q_i \leq q_{i-1} \forall i = 2, 3,\dots\}$ 
be the set of all possible fluid-scaled states.
Whenever we consider fluid limits, we assume the sequence of initial
states is such that $\qq^N(0) \to \qq^\infty \in \cS$ as $N \to \infty$.

The diffusion-scaled occupancy state is defined as
$\bar{\QQ}^N(t) = (\bar{Q}_1^N(t), \bar{Q}_2^N(t), \dots)$, with
\begin{equation}\label{eq:diffscale}
\bar{Q}_1^N(t) = - \frac{N - Q_1^N(t)}{\sqrt{{N}}}, \qquad
\bar{Q}_i^N(t) = \frac{Q_i^N(t)}{\sqrt{{N}}}, \quad i = 2,3, \dots.
\end{equation}
Note that $-\bar{Q}_1^N(t)$ corresponds to the number of vacant servers,
normalized by $\sqrt{N}$.
The reason why $Q_1^N(t)$ is centered around~$N$ while $Q_i^N(t)$,
$i = 2,3, \dots$, are not, is because for the scalable LBAs that we
pursue, the fraction of servers with exactly one task tends to one,
whereas the fraction of servers with two or more tasks tends to zero
as $N \to \infty$.

\subsection{Fluid limit for JSQ(d) policies}

We first consider the fluid limit for JSQ($d$) policies
with an arbitrary but fixed value of~$d$ as characterized
by Mitzenmacher~\cite{Mitzenmacher01} and Vvedenskaya
{\em et al.}~\cite{VDK96}.

{\em The sequence of processes $\{\qq^N(t)\}_{t \geq 0}$ has a weak
limit $\{\qq(t)\}_{t \geq 0}$ that satisfies the system
of differential equations}
\begin{equation}
\label{fluid:standard}
\frac{\dif q_i(t)}{\dif t} =
\lambda [(q_{i-1}(t))^d - (q_i(t))^d] - [q_i(t) - q_{i+1}(t)],
\quad i = 1, 2, \dots.
\end{equation}
The fluid-limit equations may be interpreted as follows.
The first term represents the rate of increase in the fraction
of servers with $i$ or more tasks due to arriving tasks that are
assigned to a server with exactly $i - 1$ tasks.
Note that the latter occurs in fluid state $\qq \in \cS$ with probability
$q_{i-1}^d - q_i^d$, i.e., the probability that all $d$~sampled servers
have $i - 1$ or more tasks, but not all of them have $i$ or more tasks.
The second term corresponds to the rate of decrease in the fraction
of servers with $i$ or more tasks due to service completions from servers
with exactly $i$ tasks, and the latter rate is given by $q_i - q_{i+1}$.

The unique fixed point of~\eqref{fluid:standard} for any $d \geq 2$ is
obtained as
\begin{equation}
\label{eq:fixedpoint1}
q_i^\star = \lambda^{\frac{d^i-1}{d-1}},
\quad i = 1, 2, \dots.
\end{equation}
It can be shown that the fixed point is asymptotically stable in the
sense that $\qq(t) \to \qq^\star$ as $t \to \infty$ for any initial fluid
state $\qq^\infty$ with $\sum_{i = 1}^{\infty} q_i^\infty < \infty$.
The fixed point reveals that the stationary queue length distribution at
each individual server exhibits super-exponential decay as $N \to \infty$,
as opposed to exponential decay for a random assignment policy.
It is worth observing that this involves an interchange of the
many-server ($N \to \infty$) and stationary ($t \to \infty$) limits.
The justification is provided by the asymptotic stability of the fixed
point along with a few further technical conditions.

\subsection{Fluid limit for JSQ policy}
\label{ssec:jsqfluid}

We now turn to the fluid limit for the ordinary JSQ policy,
which rather surprisingly was not rigorously established until fairly
recently in~\cite{MBLW16-3}, leveraging martingale functional limit
theorems and time-scale separation arguments~\cite{HK94}.

In order to state the fluid limit starting from an arbitrary
fluid-scaled occupancy state, we first introduce some additional notation.
For any fluid state $\qq \in \cS$,
denote by $m(\qq) = \min\{i: q_{i + 1} < 1\}$ the minimum queue length
among all servers.
Now if $m(\qq)=0$, then define $p_0(m(\qq))=1$ and $p_i(m(\qq))=0$
for all $i=1,2,\ldots$. 
Otherwise, in case $m(\qq)>0$, define
\begin{equation}
\label{eq:fluid-gen}
p_i(\qq) =
\begin{cases}
\min\big\{(1 - q_{m(\qq) + 1})/\lambda,1\big\} & \quad\mbox{ for }\quad i=m(\qq)-1, \\
1 - p_{m(\qq) - 1}(\qq) & \quad\mbox{ for }\quad i=m(\qq), 
\end{cases}
\end{equation}
and $p_i(\qq)=0$ otherwise.
The coefficient $p_i(\qq)$ represents the instantaneous fraction
of incoming tasks assigned to servers with a queue length of exactly~$i$
in the fluid state $\qq \in \mathcal{S}$.

{\em Any weak limit of the sequence of processes $\{\qq^N(t)\}_{t \geq 0}$
is given by the deterministic system $\{\qq(t)\}_{t \geq 0}$ satisfying
the following system of differential equations}
\begin{equation}
\label{eq:fluid}
\frac{\dif^+ q_i(t)}{\dif t} =
\lambda p_{i-1}(\qq(t)) - (q_i(t) - q_{i+1}(t)),
\quad i = 1, 2, \dots,
\end{equation}
{\em where $\dif^+/\dif t$ denotes the right-derivative.}

The unique fixed point
$\qq^\star = (q_1^\star,q_2^\star,\ldots)$ of the dynamical system
in~\eqref{eq:fluid} is given by
\begin{equation}
\label{eq:fpjsq}
q_i^\star = \left\{\begin{array}{ll} \lambda, & i = 1, \\
0, & i = 2, 3,\dots. \end{array} \right.
\end{equation}
The fixed point in~\eqref{eq:fpjsq}, in conjunction with an interchange
of limits argument, indicates that in stationarity the fraction
of servers with a queue length of two or larger under the JSQ policy
is negligible as $N \to \infty$.

\subsection{Diffusion limit for JSQ policy}
\label{ssec:diffjsq}

We next describe the diffusion limit for the JSQ policy in the
Halfin-Whitt heavy-traffic regime~\eqref{eq:HW}, as recently derived
by Eschenfeldt \& Gamarnik~\cite{EG15}.

{\em For suitable initial conditions, the sequence of processes
$\big\{\bar{\QQ}^N(t)\big\}_{t \geq 0}$ as in~\eqref{eq:diffscale} converges weakly to the limit
$\big\{\bar{\QQ}(t)\big\}_{t \geq 0}$, 
where 
$(\bar{Q}_1(t), \bar{Q}_2(t),\ldots)$
is the unique solution 
to the following system of SDEs}
\begin{equation}
\label{eq:diffusionjsq}
\begin{split}
\dif\bar{Q}_1(t) &= \sqrt{2}\dif W(t) - \beta\dif t - \bar{Q}_1(t)\dif t + \bar{Q}_2(t)\dif t-\dif U_1(t), \\
\dif\bar{Q}_2(t) &= \dif U_1(t) -  (\bar{Q}_2(t)-\bar{Q}_3(t))\dif t, \\
\dif\bQ_i(t) &= - (\bQ_i(t) - \bQ_{i+1}(t))\dif t, \quad i \geq 3,
\end{split}
\end{equation}
{\em for $t \geq 0$, where $W(\cdot)$ is the standard Brownian motion and $U_1(\cdot)$ is
the unique nondecreasing nonnegative process 
satisfying}
$\int_0^\infty \mathbbm{1}_{[\bar{Q}_1(t) < 0]} \dif U_1(t) = 0$.

The above diffusion limit implies that the mean waiting time under the
JSQ policy is of a similar order $O(1 / \sqrt{N})$ as in the
corresponding centralized M/M/$N$ queue.
Hence, we conclude that despite the distributed queueing operation
a suitable load balancing policy can deliver a similar combination
of excellent service quality and high resource utilization in the
Halfin-Whitt regime~\eqref{eq:HW} as in a centralized queueing arrangement.
It it important though to observe a subtle but fundamental difference
in the distributional properties due to the distributed versus
centralized queueing operation.
In the ordinary M/M/$N$ queue a fraction $\Pi_W^\star(\beta)$ of the
customers incur a non-zero waiting time as $N \to \infty$, but a non-zero
waiting time is only of length $1 / (\beta \sqrt{N})$ in expectation.
In contrast, under the JSQ policy, the fraction of tasks that experience
a non-zero waiting time is only of the order $O(1 / \sqrt{N})$.
However, such tasks will have to wait for the duration of a residual
service time, yielding a waiting time of the order $O(1)$.

\subsection{Heavy-traffic limits for JSQ(d) policies}

Finally, we briefly discuss the behavior of JSQ($d$) policies
for fixed~$d$ in a heavy-traffic regime
where $(N - \lambda(N)) / \eta(N) \to \beta > 0$ as $N \to \infty$
with $\eta(N)$ a positive function diverging to infinity.
Note that the case $\eta(N)= \sqrt{N}$ corresponds to the Halfin-Whitt
heavy-traffic regime~\eqref{eq:HW}.
While a complete characterization of the occupancy process
for fixed~$d$ has remained elusive so far, significant partial results
were recently obtained by Eschenfeldt \& Gamarnik~\cite{EG16}.
In order to describe the transient asymptotics, we introduce the
following rescaled processes $\bQ_i^N(t) := (N-Q_i^N(t)) / \eta(N)$,
$i = 1, 2, \ldots$.

Then, {\em for suitable initial states, on any finite time interval,
$\{\bar{\QQ}^N(t)\}_{t \geq 0}$ converges weakly to a deterministic system
$\{\bar{\QQ}(t)\}_{t \geq 0}$ that satisfies the following system of ODEs}
\begin{equation}
\frac{\dif \bQ_i(t)}{\dif t} =
- d [\bQ_i(t) - \bQ_{i-1}(t)] - [\bQ_i(t) - \bQ_{i+1}(t)],
\quad i = 1, 2, \ldots,
\end{equation}
{\em with the convention that $\bQ_0(t) \equiv 0$.}

It is noteworthy that the scaled occupancy process loses its
diffusive behavior for fixed~$d$.
It is further shown in~\cite{EG16} that with high probability the
steady-state fraction of queues with length at least
$\log_d(N/\eta(N)) - \omega(1)$ tasks approaches unity,
which in turn implies that with high probability the steady-state delay
is {\em at least} $\log_d(N/\eta(N)) - O(1)$ as $N \to \infty$.
The diffusion approximation of the JSQ($d$) policy in the
Halfin-Whitt regime~\eqref{eq:HW}, starting from a different initial
scaling, has been studied by Budhiraja \& Friedlander~\cite{BF17}.
Recently, Ying~\cite{Ying17} introduced a broad framework involving
Stein's method to analyze the rate of convergence of the scaled
steady-state occupancy process of the JSQ($2$) policy when
$\eta(N) = N^\alpha$ with $\alpha>0.8$.
The results in~\cite{Ying17} establish that in steady state,
most of the queues are of size $\log_2(N/\eta(N))+O(1),$ and thus
the steady-state delay is of order $\log_2(N/\eta(N))$.

\subsection{Universality properties}
\label{univ}


We now further explore the trade-off between delay performance
and communication overhead as a function of the diversity parameter~$d$,
in conjunction with the relative load.
The latter trade-off will be examined in an asymptotic regime where
not only the total task arrival rate $\lambda(N)$ grows with~$N$,
but also the diversity parameter depends on~$N$,
and we write $d(N)$, to explicitly reflect that.
We will specifically investigate what growth rate of $d(N)$ is required,
depending on the scaling behavior of $\lambda(N)$,
in order to asymptotically match the optimal performance of the JSQ
policy and achieve a zero mean waiting time in the limit.
The results presented in this subsection are based on~\cite{MBLW16-3},
unless specified otherwise.

\begin{theorem}{\normalfont (Universality fluid limit for JSQ($d(N)$))}
\label{fluidjsqd}
If $d(N)\to\infty$ as $N\to\infty$, then the fluid limit of the
JSQ$(d(N))$ scheme coincides with that of the ordinary JSQ policy
given by the dynamical system in~\eqref{eq:fluid}.
Consequently, the stationary occupancy states converge to the unique
fixed point in~\eqref{eq:fpjsq}.
\end{theorem}

\begin{theorem}{\normalfont (Universality diffusion limit for JSQ($d(N)$))}
\label{diffusionjsqd}
If $d(N) /( \sqrt{N} \log N)\to\infty$, then for suitable initial
conditions the weak limit of the sequence of processes
$\big\{\bar{\QQ}^{ d(N)}(t)\big\}_{t \geq 0}$ coincides with that
of the ordinary JSQ policy, and in particular, is given by the system
of SDEs in~\eqref{eq:diffusionjsq}.
\end{theorem}

The above universality properties indicate that the JSQ overhead can
be lowered by almost a factor O($N$) and O($\sqrt{N} / \log N$)
while retaining fluid- and diffusion-level optimality, respectively.
In other words, Theorems~\ref{fluidjsqd} and~\ref{diffusionjsqd}
thus reveal that it is sufficient for $d(N)$ to grow at any rate
and faster than $\sqrt{N} \log N$ in order to observe similar scaling
benefits as in a corresponding centralized M/M/$N$ queue on fluid
scale and diffusion scale, respectively.
The stated conditions are in fact close to necessary, in the sense
that if $d(N)$ is uniformly bounded and $d(N) /( \sqrt{N} \log N) \to 0$
as $N \to \infty$, then the fluid-limit and diffusion-limit paths
of the system occupancy process under the JSQ($d(N)$) scheme differ
from those under the ordinary JSQ policy, respectively.
In particular, if $d(N)$ is uniformly bounded, the mean steady-state
delay does not vanish asymptotically as $N \to \infty$.

\paragraph{High-level proof idea.}

The proofs of both Theorems~\ref{fluidjsqd} and~\ref{diffusionjsqd}
rely on a stochastic coupling construction to bound the difference
in the queue length processes between the JSQ policy and a scheme
with an arbitrary value of $d(N)$.  
This S-coupling (`S' stands for server-based) is then exploited to obtain
the fluid and diffusion limits of the JSQ($d(N)$) policy under the
conditions stated in Theorems~\ref{fluidjsqd} and~\ref{diffusionjsqd}.

\begin{figure}
\begin{center}
\begin{subfigure}{.5\textwidth}
  \centering
  \includegraphics[scale=1]{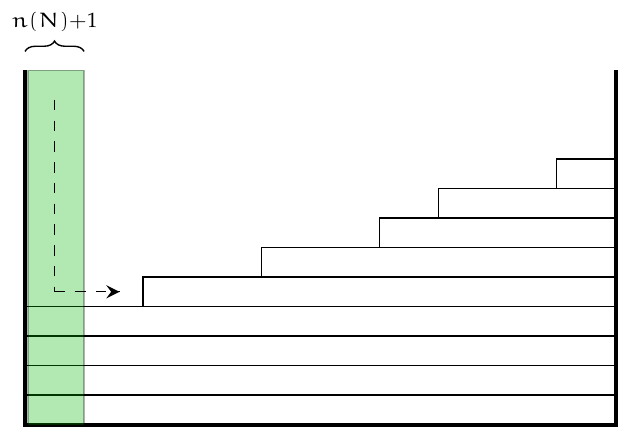}
  \caption{$\CJSQ(n(N))$ scheme\vspace{16pt}}
  \label{fig:sfigCJSQ}
\end{subfigure}%
\begin{subfigure}{.5\textwidth}
  \centering
  \includegraphics[scale=1]{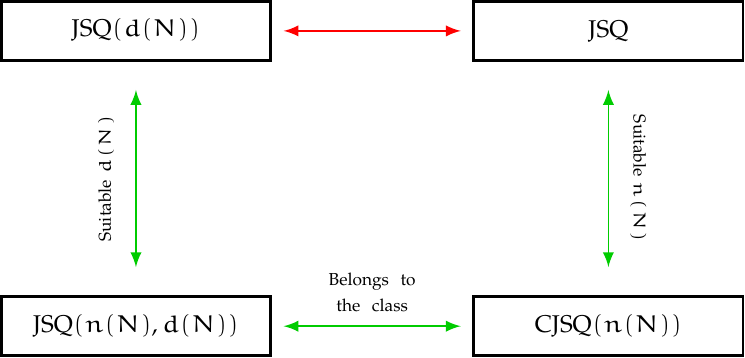}
  \caption{Asymptotic equivalence relations}
  \label{fig:sfigRelation}
\end{subfigure}
\caption{(a) High-level view of the $\CJSQ(n(N))$ class of schemes, where as in Figure~\ref{figB}, the servers are arranged in nondecreasing order of their queue lengths, and the arrival must be assigned through the left tunnel. (b) The equivalence structure is depicted for various intermediate load balancing schemes to facilitate the comparison between the JSQ$(d(N))$ scheme and the ordinary JSQ policy.
}
\label{fig:strategy}
\end{center}
\end{figure}

A direct comparison between the JSQ$(d(N))$ scheme and the ordinary JSQ policy is not straightforward, which is why the $\CJSQ(n(N))$ class of schemes is introduced as an intermediate scenario to establish the universality result.
Just like the JSQ$(d(N))$ scheme, the schemes in the class $\CJSQ(n(N))$ may be thought of as ``sloppy'' versions of the JSQ policy, in the sense that tasks are not necessarily assigned to a server with the shortest queue length but to one of the $n(N)+1$
lowest ordered servers, as graphically illustrated in Figure~\ref{fig:sfigCJSQ}.
In particular, for $n(N)=0$, the class only includes the ordinary JSQ policy. 
Note that the JSQ$(d(N))$ scheme is guaranteed to identify the lowest ordered server, but only among a randomly sampled subset of $d(N)$ servers.
In contrast, a scheme in the $\CJSQ(n(N))$ class only guarantees that
one of the $n(N)+1$ lowest ordered servers is selected, but 
across the entire pool of $N$ servers. 
It may be shown that for sufficiently small $n(N)$, any scheme from the class $\CJSQ(n(N))$ is still `close' to the ordinary JSQ policy. 
It can further be proved that for sufficiently large $d(N)$ relative to $n(N)$ we can construct a scheme
called JSQ$(n(N),d(N))$, belonging to the $\CJSQ(n(N))$ class, which differs `negligibly' from the JSQ$(d(N))$ scheme. 
Therefore,  for a `suitable' choice of $d(N)$ the idea is to produce a `suitable' $n(N)$.
This proof strategy is schematically represented in Figure~\ref{fig:sfigRelation}.

In order to prove the stochastic comparisons among the various schemes,
the many-server system is described as an ensemble of stacks,
in a way that two different ensembles can be ordered.
This stack formulation has also been considered in the literature
for establishing the stochastic optimality properties of the JSQ
policy \cite{towsley,Towsley95,Towsley1992}.
However, it is only through the stack arguments developed
in~\cite{MBLW16-3} that the comparison results can be extended to
any scheme from the class CJSQ.

\section{Blocking and infinite-server dynamics}
\label{bloc}

The basic scenario that we have focused on so far involved
single-server queues.
In this section we turn attention to a system with parallel server
pools, each with $B$~servers, where $B$ can possibly be infinite.
As before, tasks must immediately be forwarded to one of the server pools,
but also directly start execution or be discarded otherwise.
The execution times are assumed to be exponentially distributed,
and do not depend on the number of other tasks receiving service
simultaneously.
The current scenario will be referred to as `infinite-server dynamics',
in contrast to the earlier single-server queueing dynamics.

As it turns out, the JSQ policy has similar stochastic optimality
properties as in the case of single-server queues, and in particular
stochastically minimizes the cumulative number of discarded tasks
\cite{STC93,J89,M87,MS91}.
However, the JSQ policy also suffers from a similar scalability issue
due to the excessive communication overhead in large-scale systems,
which can be mitigated through JSQ($d$) policies.
Results of Turner~\cite{T98} and recent papers by Karthik
{\em et al.}~\cite{KMM17}, Mukhopadhyay {\em et al.}~\cite{MKMG15,MMG15}, 
and Xie {\em et al.}~\cite{XDLS15} indicate that JSQ($d$) policies
provide similar ``power-of-choice'' gains for loss probabilities.
It may be shown though that the optimal performance of the JSQ policy
cannot be matched for any fixed value of~$d$.

Motivated by these observations, we explore the trade-off between
performance and communication overhead for infinite-server dynamics.
We will demonstrate that the optimal performance of the JSQ policy
can be asymptotically retained while drastically reducing the
communication burden, mirroring the universality properties described
in Section~\ref{univ} for single-server queues.
The results presented in the remainder of the section are extracted
from~\cite{MBLW16-4}, unless indicated otherwise.

\subsection{Fluid limit for JSQ policy}
\label{ssec:jsqfluid-infinite}

As in Subsection~\ref{ssec:jsqfluid}, for any fluid state $\qq \in \cS$,
denote by $m(\qq) = \min\{i: q_{i + 1} < 1\}$ the minimum queue length
among all servers.
Now if $m(\qq)=0$, then define $p_0(m(\qq))=1$ and $p_i(m(\qq))=0$
for all $i=1,2,\ldots$. 
Otherwise, in case $m(\qq)>0$, define
\begin{equation}
\label{eq:fluid-prob-infinite}
p_{i}(\qq) =
\begin{cases}
\min\big\{m(\qq)(1 - q_{m(\qq) + 1})/\lambda,1\big\} & \quad \mbox{ for }
\quad i=m(\qq)-1, \\
1 - p_{ m(\qq) - 1}(\qq) & \quad \mbox{ for } \quad i=m(\qq), 
\end{cases}
\end{equation}
and $p_i(\qq)=0$ otherwise.
As before, the coefficient $p_i(\qq)$ represents the instantaneous
fraction of incoming tasks assigned to servers with a queue length
of exactly~$i$ in the fluid state $\qq \in \mathcal{S}$.

{\em Any weak limit of the sequence of processes $\{\qq^N(t)\}_{t \geq 0}$
is given by the deterministic system $\{\qq(t)\}_{t \geq 0}$ satisfying
the following of differential equations}
\begin{equation}
\label{eq:fluid-infinite}
\frac{\dif^+ q_i(t)}{\dif t} =
\lambda p_{i-1}(\qq(t)) - i (q_i(t) - q_{i+1}(t)),
\quad i = 1, 2, \dots,
\end{equation}
{\em where $\dif^+/\dif t$ denotes the right-derivative.}

Equations~\eqref{eq:fluid-prob-infinite} and \eqref{eq:fluid-infinite}
are to be contrasted with Equations~\eqref{eq:fluid-gen}
and~\eqref{eq:fluid}.
While the form of~\eqref{eq:fluid-prob-infinite} and the
evolution equations~\eqref{eq:fluid-infinite} of the limiting
dynamical system remains similar to that of~\eqref{eq:fluid-gen}
and~\eqref{eq:fluid}, respectively, an additional factor $m(\qq)$
appears in~\eqref{eq:fluid-prob-infinite} and the rate of decrease
in~\eqref{eq:fluid-infinite} now becomes $i (q_i - q_{i+1})$,
reflecting the infinite-server dynamics.

Let $K := \lfloor \lambda \rfloor$ and $f := \lambda - K$ denote the
integral and fractional parts of~$\lambda$, respectively.
It is easily verified that, assuming $\lambda<B$, the unique fixed point
of the dynamical system in~\eqref{eq:fluid-infinite} is given by
\begin{equation}
\label{eq:fixed-point-infinite}
q_i^\star = \left\{\begin{array}{ll} 1 & i = 1, \dots, K \\
f & i = K + 1 \\
0 & i = K + 2, \dots, B, \end{array} \right.
\end{equation}
and thus $\sum_{i=1}^{B} q_i^\star = \lambda$.
This is consistent with the results in Mukhopadhyay
{\em et al.}~\cite{MKMG15,MMG15} and Xie {\em et al.}~\cite{XDLS15}
for fixed~$d$, where taking $d \to \infty$ yields the same fixed point.
The fixed point in~\eqref{eq:fixed-point-infinite}, in conjunction
with an interchange of limits argument, indicates that in stationarity
the fraction of server pools with at least $K+2$ and at most $K-1$
active tasks is negligible as $N \to \infty$.

\subsection{Diffusion limit for JSQ policy}
\label{ssec:jsq-diffusion-infinite}

As it turns out, the diffusion-limit results may be qualitatively
different, depending on whether $f = 0$ or $f > 0$,
and we will distinguish between these two cases accordingly.
Observe that for any assignment scheme, in the absence of overflow events,
the total number of active tasks evolves as the number of jobs
in an M/M/$\infty$ system, for which the diffusion limit is well-known.
For the JSQ policy, it can be established that the total number
of server pools with $K - 2$ or less and $K + 2$ or more tasks is
negligible on the diffusion scale.
If $f > 0$, the number of server pools with $K - 1$ tasks is negligible
as well, and the dynamics of the number of server pools with $K$
or $K + 1$ tasks can then be derived from the known diffusion limit
of the total number of tasks mentioned above.
In contrast, if $f = 0$, the number of server pools with $K - 1$ tasks
is not negligible on the diffusion scale, and the limiting behavior is
qualitatively different, but can still be characterized.
We refer to~\cite{MBLW16-4} for further details.

\subsection{Universality of JSQ(d) policies in infinite-server scenario}
\label{ssec:univ-infinite}

As in Subsection~\ref{univ}, we now further explore the trade-off
between performance and communication overhead as a function
of the diversity parameter~$d(N)$, in conjunction with the relative load.
We will specifically investigate what growth rate of $d(N)$ is required,
depending on the scaling behavior of $\lambda(N)$, in order to
asymptotically match the optimal performance of the JSQ policy.

\begin{theorem}{\normalfont (Universality fluid limit for JSQ($d(N)$))}
\label{fluidjsqd-infinite}
If $d(N)\to\infty$ as $N\to\infty$, then the fluid limit of the
JSQ$(d(N))$ scheme coincides with that of the ordinary JSQ policy
given by the dynamical system in~\eqref{eq:fluid-infinite}. 
Consequently, the stationary occupancy states converge to the unique
fixed point in~\eqref{eq:fixed-point-infinite}.
\end{theorem}

In order to state the universality result on diffusion scale,
define in case $f > 0$, $f(N):=\lambda(N)-K(N)$, \small
\[
\bar{Q}_i^{d(N)}(t) := \dfrac{N - Q_i^{d(N)}(t)}{\sqrt{N}} \: (i \leq K),
\ 
\bar{Q}_{K+1}^{d(N)}(t) := \dfrac{Q_{K+1}^{d(N)}(t) - f(N)}{\sqrt{N}},
\ 
\bar{Q}_i^{d(N)}(t) := \frac{Q_i^{d(N)}(t)}{\sqrt{N}}\geq 0 \: (i \geq K + 2),
\]\normalsize
and otherwise, if $f = 0$, assume $(KN-\lambda(N))/\sqrt{N}\to\beta\in \R$ as $N\to\infty$, and define \small
\[\footnotesize
\hQ_{K-1}^{d(N)}(t) :=
\sum_{i=1}^{K-1} \dfrac{N - Q_i^{d(N)}(t)}{\sqrt{N}}, \ 
\hQ_K^{d(N)}(t) := \dfrac{N - Q_K^{d(N)}(t)}{\sqrt{N}}, \ 
\hQ_i^{d(N)}(t) := \dfrac{Q_i^{d(N)}(t)}{\sqrt{N}} \geq 0 \: (i \geq K + 1).
\]\normalsize

\begin{theorem}[\normalfont Universality diffusion limit for JSQ($d(N)$)]
\label{diffusionjsqd-infinite}
Assume $d(N) / (\sqrt{N} \log N) \to \infty$. 
Under suitable initial conditions \\
{\normalfont (i)}
If $f>0$, then $\bQ_i^{d(N)}(\cdot)$ converges to the zero process for $i\neq K+1$, and $\bQ^{d(N)}_{K+1}(\cdot)$ converges weakly to the Ornstein-Uhlenbeck process satisfying the SDE
$d\bar{Q}_{K+1}(t)=-\bar{Q}_{K+1}(t)dt+\sqrt{2\lambda}dW(t)$,
where $W(\cdot)$ is the standard Brownian motion. \\
{\normalfont (ii)}
If $f=0$, then $\hQ_{K-1}^{d(N)}(\cdot)$ converges weakly to the zero process, and $(\hQ_{K}^{d(N)}(\cdot), \hQ_{K+1}^{d(N)}(\cdot))$ converges weakly to $(\hQ_{K}(\cdot), \hQ_{K+1}(\cdot))$, described by the unique solution of the following system of SDEs:
\begin{align*}
\dif\hQ_{K}(t) &= \sqrt{2 K} \dif W(t) - (\hQ_K(t) + K \hQ_{K+1}(t))\dif t +
\beta \dif t + \dif V_1(t) \\
\dif\hQ_{K+1}(t) &= \dif V_1(t) - (K + 1) \hQ_{K+1}(t)\dif t,
\end{align*}
where $W(\cdot)$ is the standard Brownian motion, and $V_1(\cdot)$ is the unique
nondecreasing process satisfying
$\int_0^t \ind{\hQ_K(s)\geq 0} \dif V_1(s) = 0$.
\end{theorem}

Given the asymptotic results for the JSQ policy
in Subsections~\ref{ssec:jsqfluid-infinite}
and~\ref{ssec:jsq-diffusion-infinite}, the proofs of the asymptotic
results for the JSQ$(d(N))$ scheme in Theorems~\ref{fluidjsqd-infinite}
and~\ref{diffusionjsqd-infinite} involve establishing a universality
result which shows that the limiting processes for the JSQ$(d(N))$
scheme are `$g(N)$-alike' to those for the ordinary JSQ
policy for suitably large~$d(N)$.
Loosely speaking, if two schemes are $g(N)$-alike, then in some sense,
the associated system occupancy states are indistinguishable
on $g(N)$-scale.

The next theorem states a sufficient criterion for the JSQ$(d(N))$
scheme and the ordinary JSQ policy to be $g(N)$-alike, and thus,
provides the key vehicle in establishing the universality result.

\begin{theorem}
\label{th:pwr of d}
Let $g: \N \to \R_+$ be a function diverging to infinity.
Then the JSQ policy and the JSQ$(d(N))$ scheme are $g(N)$-alike,
with $g(N) \leq N$, if \\
{\rm (i)} $d(N) \to \infty$ for $g(N) = O(N)$, 
{\rm (ii)} $d(N) \left(\frac{N}{g(N)}\log\left(\frac{N}{g(N)}\right)\right)^{-1} \to \infty$ for $g(N)=o(N)$.
\end{theorem}

The proof of Theorem~\ref{th:pwr of d} relies on a novel coupling
construction, called T-coupling (`T' stands for task-based),
which will be used to (lower and upper) bound the difference
of occupancy states of two arbitrary schemes.
This T-coupling~\cite{MBLW16-4} is distinct from and inherently
stronger than the S-coupling used in Subsection~\ref{univ} in the
single-server queueing scenario.
Note that in the current infinite-server scenario, the departures
of the ordered server pools cannot be coupled, mainly since the
departure rate at the $m^{\rm th}$ ordered server pool,
for some $m = 1, 2, \ldots, N$, depends on its number of active tasks.
The T-coupling is also fundamentally different from the coupling
constructions used in establishing the weak majorization results
in \cite{Winston77,towsley,Towsley95,Towsley1992,W78} in the context
of the ordinary JSQ policy in the single-server queueing scenario,
and in \cite{STC93,J89,M87,MS91} in the scenario of state-dependent
service rates.

\section{Universality of load balancing in networks}
\label{networks}

In this section we return to the single-server queueing dynamics,
and extend the universality properties to network scenarios,
where the $N$~servers are assumed to be inter-connected by some
underlying graph topology~$G_N$.
Tasks arrive at the various servers as independent Poisson processes
of rate~$\lambda$, and each incoming task is assigned to whichever
server has the smallest number of tasks among the one where it arrives
and its neighbors in~$G_N$.  
Thus, in case $G_N$ is a clique, each incoming task is assigned to the
server with the shortest queue across the entire system,
and the behavior is equivalent to that under the JSQ policy.
The stochastic optimality properties of the JSQ policy thus imply that
the queue length process in a clique will be better balanced
and smaller (in a majorization sense) than in an arbitrary graph~$G_N$.

Besides the prohibitive communication overhead discussed earlier,
a further scalability issue of the JSQ policy arises when executing
a task involves the use of some data.
Storing such data for all possible tasks on all servers will typically
require an excessive amount of storage capacity.
These two burdens can be effectively mitigated in sparser graph
topologies where tasks that arrive at a specific server~$i$ are only
allowed to be forwarded to a subset of the servers ${\mathcal N}_i$.
For the tasks that arrive at server~$i$, queue length information
then only needs to be obtained from servers in ${\mathcal N}_i$,
and it suffices to store replicas of the required data on the
servers in ${\mathcal N}_i$.
The subset ${\mathcal N}_i$ containing the peers of server~$i$ can
be naturally viewed as its neighbors in some graph topology~$G_N$.
In this section we focus on the results in~\cite{MBL17}
for the case of undirected graphs,
but most of the analysis can be extended to directed graphs.

The above model has been studied in~\cite{G15,T98}, focusing
on certain fixed-degree graphs and in particular ring topologies.
The results demonstrate that the flexibility to forward tasks
to a few neighbors, or even just one, with possibly shorter queues
significantly improves the performance in terms of the waiting time
and tail distribution of the queue length.
This resembles the ``power-of-choice'' gains observed for JSQ($d$)
policies in complete graphs.
However, the results in \cite{G15,T98} also establish that
the performance sensitively depends on the underlying graph topology,
and that selecting from a fixed set of $d - 1$ neighbors typically
does not match the performance of re-sampling $d - 1$ alternate
servers for each incoming task from the entire population,
as in the power-of-$d$ scheme in a complete graph.

If tasks do not get served and never depart but simply accumulate,
then the scenario described above amounts to a so-called balls-and-bins
problem on a graph.
Viewed from that angle, a close counterpart of our setup is studied
in Kenthapadi \& Panigrahy~\cite{KP06}, where in our terminology each
arriving task is routed to the shortest of $d \geq 2$ randomly
selected neighboring queues.

The key challenge in the analysis of load balancing on arbitrary graph
topologies is that one needs to keep track
of the evolution of number of tasks at each vertex along with their
corresponding neighborhood relationship.
This creates a major problem in constructing a tractable Markovian
state descriptor, and renders a direct analysis of such processes
highly intractable.
Consequently, even asymptotic results for load balancing processes
on an arbitrary graph have remained scarce so far.
The approach in~\cite{MBL17} is radically different, and aims
at comparing the load balancing process on an arbitrary graph
with that on a clique.
Specifically, rather than analyzing the behavior for a given class
of graphs or degree value, the analysis explores for what types
of topologies and degree properties the performance is asymptotically
similar to that in a clique.
The proof arguments in~\cite{MBL17} build on the stochastic coupling
constructions developed in Subsection~\ref{univ} for JSQ($d$) policies.
Specifically, the load balancing process on an arbitrary graph is
viewed as a `sloppy' version of that on a clique, and several other
intermediate sloppy versions are constructed.

Let $Q_i(G_N, t)$ denote the number of servers with queue length
at least~$i$ at time~$t$, $i = 1, 2, \ldots$, and let the
fluid-scaled variables $q_i(G_N, t) := Q_i(G_N, t) / N$ be the
corresponding fractions.
Also, in the Halfin-Whitt heavy-traffic regime~\eqref{eq:HW},
define the centered and diffusion-scaled variables
$\bar{Q}_1(G_N,t) := - (N - Q_1(G_N,t)) / \sqrt{N}$
and $\bar{Q}_i(G_N,t) := Q_i(G_N,t) / \sqrt{N}$ for $i = 2, 3, \ldots$,
analogous to~\eqref{eq:diffscale}. \\

The next definition introduces two notions of \emph{asymptotic optimality}.

\begin{definition}[{Asymptotic optimality}]
\label{def:opt}
A graph sequence $\GG = \{G_N\}_{N \geq 1}$ is called `asymptotically
optimal on $N$-scale' or `$N$-optimal', if for any $\lambda < 1$,
the scaled occupancy process $(q_1(G_N, \cdot), q_2(G_N, \cdot), \ldots)$
converges weakly, on any finite time interval, to the process
$(q_1(\cdot), q_2(\cdot),\ldots)$ given by~\eqref{eq:fluid}.

Moreover, a graph sequence $\GG = \{G_N\}_{N \geq 1}$ is called
`asymptotically optimal on $\sqrt{N}$-scale' or `$\sqrt{N}$-optimal',
if in the Halfin-Whitt heavy-traffic regime~\eqref{eq:HW},
on any finite time interval, the process
$(\bQ_1(G_N, \cdot), \bQ_2(G_N, \cdot), \ldots)$
converges weakly to the process $(\bQ_1(\cdot), \bQ_2(\cdot), \ldots)$
given by~\eqref{eq:diffusionjsq}.
\end{definition}

Intuitively speaking, if a graph sequence is $N$-optimal
or $\sqrt{N}$-optimal, then in some sense, the associated occupancy
processes are indistinguishable from those of the sequence of cliques
on $N$-scale or $\sqrt{N}$-scale.
In other words, on any finite time interval their occupancy processes
can differ from those in cliques by at most $o(N)$ or $o(\sqrt{N})$,
respectively. 

\subsection{Asymptotic optimality criteria for deterministic graph sequences}

We now develop a criterion for asymptotic optimality of an arbitrary
deterministic graph sequence on different scales.
We first introduce some useful notation, and two measures
of \emph{well-connectedness}.
Let $G = (V, E)$ be any graph.
For a subset $U \subseteq V$, define $\com(U) := |V\setminus N[U]|$
to be the set of all vertices that are disjoint from $U$, where
$N[U] := U\cup \{v \in V:\ \exists\ u \in U \mbox{ with } (u, v) \in E\}$.
For any fixed $\varepsilon > 0$ define
\begin{equation}
\label{def:dis}
\dis_1(G,\varepsilon) := \sup_{U\subseteq V, |U|\geq \varepsilon |V|}\com(U),
\qquad
\dis_2(G,\varepsilon) := \sup_{U\subseteq V, |U|\geq \varepsilon \sqrt{|V|}}\com(U).
\end{equation}

The next theorem provides sufficient conditions for asymptotic
optimality on $N$-scale and $\sqrt{N}$-scale in terms of the above
two well-connectedness measures.

\begin{theorem}
\label{th:det-seq}
For any graph sequence $\GG = \{G_N\}_{N \geq 1}$,
{\normalfont(i)} $\GG$ is $N$-optimal if for any $\varepsilon > 0$, 
$\dis_1(G_N, \varepsilon) / N \to 0$ as $N \to \infty$.
{\normalfont(ii)} $\GG$ is $\sqrt{N}$-optimal if for any $\varepsilon > 0$, 
$\dis_2(G_N, \varepsilon) / \sqrt{N} \to 0$ as $N \to \infty$.
\end{theorem}

The next corollary is an immediate consequence of Theorem~\ref{th:det-seq}.

\begin{corollary}
Let $\GG = \{G_N\}_{N \geq 1}$ be any graph sequence.
Then {\rm(i)} If $d_{\min}(G_N) = N - o(N)$, then $\GG$ is $N$-optimal,
and 
{\rm(ii)} If $d_{\min}(G_N) = N - o(\sqrt{N})$, then $\GG$ is
$\sqrt{N}$-optimal.
\end{corollary}

We now provide a sketch of the main proof arguments
for Theorem~\ref{th:det-seq} as used in~\cite{MBL17},
focusing on the proof of $N$-optimality. 
The proof of $\sqrt{N}$-optimality follows along similar lines.
First of all, it can be established that if a system is able to assign
each task to a server in the set $\cS^N(n(N))$ of the $n(N)$ nodes
with shortest queues, where $n(N)$ is $o(N)$, then it is $N$-optimal. 
Since the underlying graph is not a clique however (otherwise there
is nothing to prove), for any $n(N)$ not every arriving task can be
assigned to a server in $\cS^N(n(N))$.
Hence, a further stochastic comparison property is proved
in~\cite{MBL17} implying that if on any finite time interval
of length~$t$, the number of tasks $\Delta^N(t)$ that are not assigned
to a server in $\cS^N(n(N))$ is $o_P(N)$, then the system is
$N$-optimal as well.
The $N$-optimality can then be concluded when $\Delta^N(t)$ is $o_P(N)$,
which is demonstrated in \cite{MBL17} under the condition
that $\dis_1(G_N, \varepsilon) / N \to 0$ as $N \to \infty$ as stated
in Theorem~\ref{th:det-seq}.

\subsection{Asymptotic optimality of random graph sequences}

Next we investigate how the load balancing process behaves on random
graph topologies. 
Specifically, we aim to understand what types of graphs are
asymptotically optimal in the presence of randomness
(i.e., in an average-case sense).
Theorem~\ref{th:inhom} below establishes sufficient conditions
for asymptotic optimality of a sequence of inhomogeneous random graphs.
Recall that a graph $G' = (V', E')$ is called a supergraph
of $G = (V, E)$ if $V = V'$ and $E \subseteq E'$.

\begin{theorem}
\label{th:inhom}
Let $\GG= \{G_N\}_{N \geq 1}$ be a graph sequence such that for each~$N$,
$G_N = (V_N, E_N)$ is a super-graph of the inhomogeneous random graph
$G_N'$ where any two vertices $u, v \in V_N$ share an edge
with probability $p_{uv}^N$.
\begin{enumerate}[{\normalfont (i)}]
\item If for each $\varepsilon>0$, there exists subsets of vertices $V_N^\varepsilon\subseteq V_N$ with $|V_N^\varepsilon|<\varepsilon N$, such that $\inf\ \{p^N_{uv}: u, v\in V_N^\varepsilon\}$ is $\omega(1/N)$, then $\GG$ is $N$-optimal.
\item If for each $\varepsilon>0$, there exists subsets of vertices $V_N^\varepsilon\subseteq V_N$ with $|V_N^\varepsilon|<\varepsilon \sqrt{N}$, such that $\inf\ \{p^N_{uv}: u, v\in V_N^\varepsilon\}$ is $\omega(\log(N)/\sqrt{N})$, then $\GG$ is $\sqrt{N}$-optimal.
\end{enumerate}
\end{theorem}

The proof of Theorem~\ref{th:inhom} relies on Theorem~\ref{th:det-seq}.
Specifically, if $G_N$ satisfies conditions~(i) and~(ii) in
Theorem~\ref{th:inhom}, then the corresponding conditions~(i) and~(ii)
in Theorem~\ref{th:det-seq} hold.

As an immediate corollary to Theorem~\ref{th:inhom} we obtain
an optimality result for the sequence of Erd\H{o}s-R\'enyi random graphs.

\begin{corollary}
\label{cor:errg}
Let $\GG = \{G_N\}_{N \geq 1}$ be a graph sequence such that for each~$N$,
$G_N$ is a super-graph of $\ER_N(p(N))$, and $d(N) = (N-1) p(N)$.
Then
{\normalfont (i)}
If $d(N) \to \infty$ as $N \to \infty$, then $\GG$ is $N$-optimal.
{\normalfont (ii)}
If $d(N) / (\sqrt{N} \log N) \to \infty$ as $N\to\infty$, then $\GG$ is
$\sqrt{N}$-optimal.
\end{corollary}

The growth rate condition for $N$-optimality
in Corollary~\ref{cor:errg}~(i) is not only sufficient,
but necessary as well.
Thus informally speaking, $N$-optimality is achieved under the minimum
condition required as long as the underlying topology is suitably random.

\section{Token-based load balancing}
\label{token}

While a zero waiting time can be achieved in the limit by sampling
only $d(N) =o(N)$ servers as Sections~\ref{univ}, \ref{bloc}
and~\ref{networks} showed, even in network scenarios, the amount
of communication overhead in terms of $d(N)$ must still grow with~$N$.
As mentioned earlier, this can be avoided by introducing memory at the
dispatcher, in particular maintaining a record of only vacant servers,
and assigning tasks to idle servers, if there are any,
or to a uniformly at random selected server otherwise.
This so-called Join-the-Idle-Queue (JIQ) scheme \cite{BB08,LXKGLG11}
can be implemented through a simple token-based mechanism generating
at most one message per task.
Remarkably enough, even with such low communication overhead,
the mean waiting time and the probability of a non-zero waiting time
vanish under the JIQ scheme in both the fluid and diffusion regimes,
as we will discuss in the next two subsections.

\subsection{Asymptotic optimality of JIQ scheme}
We first consider the fluid limit of the JIQ scheme.
Let $q_i^N(\infty)$ be a random variable denoting the
process $q_i^N(\cdot)$ in steady state.
It was proved in~\cite{Stolyar15} for the JIQ scheme (under very broad conditions),
\begin{equation}
\label{eq:fpjiq}
q_1^N(\infty) \to \lambda, \qquad q_i^N(\infty) \to 0 \quad
\mbox{ for all } i \geq 2, \qquad \mbox{ as } \quad N \to \infty.
\end{equation}
The above equation in conjunction with the PASTA property yields that
the steady-state probability of a non-zero wait vanishes as $N \to \infty$,
thus exhibiting asymptotic optimality of the JIQ scheme on fluid scale.\\

We now turn to the diffusion limit of the JIQ scheme.
\begin{theorem}{\normalfont (Diffusion limit for JIQ)}
\label{diffusionjiq}
In the Halfin-Whitt heavy-traffic regime~\eqref{eq:HW},
under suitable initial conditions, the weak limit of the sequence of centered and diffusion-scaled occupancy process in~\eqref{eq:diffscale} coincides with that of the ordinary JSQ policy given by the system of SDEs in~\eqref{eq:diffusionjsq}.
\end{theorem}

The above theorem implies that for suitable initial states,
on any finite time interval, the occupancy process under the JIQ scheme
is indistinguishable from that under the JSQ policy.
The proof of Theorem~\ref{diffusionjiq} relies on a coupling
construction as described in greater detail in~\cite{MBLW16-1}.
The idea is to compare the occupancy processes of two systems
following JIQ and JSQ policies, respectively. 
Comparing the JIQ and JSQ policies is facilitated when viewed as follows:
(i) If there is an idle server in the system, both JIQ and JSQ perform
similarly,
(ii)~Also, when there is no idle server and only $O(\sqrt{N})$ servers
with queue length two, JSQ assigns the arriving task to a server
with queue length one. 
In that case, since JIQ assigns at random, the probability that the
task will land on a server with queue length two and thus JIQ acts
differently than JSQ is $O(1/\sqrt{N})$.
Since on any finite time interval the number of times an arrival finds
all servers busy is at most $O(\sqrt{N})$, all the arrivals except
an $O(1)$ of them are assigned in exactly the same manner in both JIQ
and JSQ, which then leads to the same scaling limit for both policies.

\subsection{Multiple dispatchers}
\label{multiple}

So far we have focused on a basic scenario with a single dispatcher.
Since it is not uncommon for LBAs to operate across multiple dispatchers
though, we consider in this subsection a scenario with $N$ parallel
identical servers as before and $R \geq 1$ dispatchers.
(We will assume the number of dispatchers to remain fixed as the
number of servers grows large, but a further natural scenario would
be for the number of dispatchers $R(N)$ to scale with the number
of servers as considered by Mitzenmacher~\cite{Mitzenmacher16},
who analyzes the case $R(N) = r N$ for some constant~$r$, so that the
relative load of each dispatcher is $\lambda r$.)
Tasks arrive at dispatcher~$r$ as a Poisson process of rate
$\alpha_r \lambda N$, with $\alpha_r > 0$, $r = 1, \dots, R$,
$\sum_{r = 1}^{R} \alpha_r = 1$, and $\lambda$ denoting the task
arrival rate per server.
For conciseness, we denote $\alpha = (\alpha_1, \dots, \alpha_R)$,
and without loss of generality we assume that the dispatchers are
indexed such that $\alpha_1 \geq \alpha_2 \geq \dots \geq \alpha_R$.

When a server becomes idle, it sends a token to one of the
dispatchers selected uniformly at random, advertising its availability.
When a task arrives at a dispatcher which has tokens available,
one of the tokens is selected, and the task is immediately forwarded
to the corresponding server.

We distinguish two scenarios when a task arrives at a dispatcher
which has no tokens available, referred to as the {\em blocking\/}
and {\em queueing\/} scenario respectively.
In the blocking scenario, the incoming task is blocked
and instantly discarded.
In the queueing scenario, the arriving task is forwarded to one
of the servers selected uniformly at random.
If the selected server happens to be idle, then the outstanding
token at one of the other dispatchers is revoked.

In the queueing scenario we assume $\lambda < 1$, which is not only
necessary but also sufficient for stability.
Denote by $B(R, N, \lambda, \alpha)$ the steady-state blocking
probability of an arbitrary task in the blocking scenario.
Also, denote by $W(R, N, \lambda, \alpha)$ a random variable
with the steady-state waiting-time distribution of an arbitrary task
in the queueing scenario.

Scenarios with multiple dispatchers have received limited attention
in the literature, and the scant papers that exist
\cite{LXKGLG11,Mitzenmacher16,Stolyar17} almost exclusively assume
that the loads at the various dispatchers are strictly equal.
 In these cases the fluid limit, for suitable initial states, is the
same as that for a single dispatcher,
and in particular the fixed point is the same,
hence, the JIQ scheme continues to achieve asymptotically optimal
delay performance with minimal communication overhead.
As one of the few exceptions, \cite{BBL17a} allows the loads at the
various dispatchers to be different.

\paragraph{Results for blocking scenario.}

For the blocking scenario, it is established in \cite{BBL17a} that,
\begin{equation*}
B(R,N,\lambda,\alpha) \to \max\{1-R\alpha_R,1-1/\lambda\} \quad
\mbox{ as } N \to \infty.
\end{equation*}
This result shows that in the many-server limit the system performance
in terms of blocking is either determined by the relative load of the
least-loaded dispatcher, or by the aggregate load.
This indirectly reveals that, somewhat counter-intuitively,
it is the least-loaded dispatcher that throttles tokens and leaves
idle servers stranded, thus acting as bottleneck.

\paragraph{Results for queueing scenario.}

For the queueing scenario, it is shown in~\cite{BBL17a} that, for fixed
$\lambda < 1$
\[
\mathbb{E}[W(R, N, \lambda, \alpha)] \to
\frac{\lambda_2(R, \lambda, \alpha)}{1 - \lambda_2(R, \lambda, \alpha)}
\quad \mbox{ as } N \to \infty,
\]
where $\lambda_2(R,\lambda,\alpha) =
1 - \frac{1 - \lambda \sum_{i=1}^{r^\star} \alpha_i}{1 - \lambda r^\star / R}$,
with $r^\star = \sup\big\{r \big| \alpha_r > \frac{1}{R}
\frac{1 - \lambda \sum_{i=1}^{r}\alpha_i}{1 - \lambda r/R}\big\}$,
may be interpreted as the rate at which tasks are forwarded
to randomly selected servers. \\

When the arrival rates at all dispatchers are strictly equal, i.e.,
$\alpha_1 = \dots = \alpha_R = 1 / R$, the above results indicate that
the stationary blocking probability and the mean waiting time
asymptotically vanish as $N \to \infty$, which is in agreement with
the observations in~\cite{Stolyar17} mentioned above.
However, when the arrival rates at the various dispatchers are not
perfectly equal, so that $\alpha_R < 1 / R$, the blocking probability
and mean waiting time are strictly positive in the limit,
even for arbitrarily low overall load and an arbitrarily small degree
of skewness in the arrival rates.
Thus, the ordinary JIQ scheme fails to achieve asymptotically optimal
performance for heterogeneous dispatcher loads.

In order to counter the above-described performance degradation for
asymmetric dispatcher loads, \cite{BBL17a} proposes two enhancements.
Enhancement~A uses a non-uniform token allotment:
When a server becomes idle, it sends a token to dispatcher~$r$
with probability~$\beta_r$.
Enhancement~B involves a token exchange mechanism:
Any token is transferred to a uniformly randomly selected
dispatcher at rate~$\nu$.
Note that the token exchange mechanism only creates a constant
communication overhead per task as long as the rate~$\nu$ does not
depend on the number of servers~$N$, and thus preserves the
scalability of the basic JIQ scheme.

The above enhancements can achieve asymptotically optimal performance for
suitable values of the $\beta_r$ parameters and the exchange rate~$\nu$.
Specifically, the stationary blocking probability in the blocking
scenario and the mean waiting time in the queueing scenario
asymptotically vanish as $N \to \infty$, upon using Enhancement~A
with $\beta_r = \alpha_r$ or Enhancement~B
with $\nu \geq \frac{\lambda}{1 - \lambda}(\alpha_1 R - 1)$.

\section{Redundancy policies and alternative scaling regimes}
\label{miscellaneous}

In this section we discuss somewhat related redundancy policies
and alternative scaling regimes and performance metrics.

\paragraph{Redundancy-d policies.}

So-called redundancy-$d$ policies involve a somewhat similar operation
as JSQ($d$) policies, and also share the primary objective of ensuring
low delays \cite{AGSS13,VGMSRS13}.
In a redundancy-$d$ policy, $d \geq 2$ candidate servers are selected
uniformly at random (with or without replacement) for each arriving task,
just like in a JSQ($d$) policy.
Rather than forwarding the task to the server with the shortest queue
however, replicas are dispatched to all sampled servers.

Two common options can be distinguished for abortion of redundant clones.
In the first variant, as soon as the first replica starts service,
the other clones are abandoned.
In this case, a task gets executed by the server which had the smallest
workload at the time of arrival (and which may or may not have had the
shortest queue length) among the sampled servers.
This may be interpreted as a power-of-$d$ version of the Join-the-Smallest
Workload (JSW) policy discussed in Subsection~\ref{ssec:jsw}.
In the second option the other clones of the task are not aborted
until the first replica has completed service (which may or may not
have been the first replica to start service).
While a task is only handled by one of the servers in the former case, 
it may be processed by several servers in the latter case.

\paragraph{Conventional heavy traffic.}

It is also worth mentioning some asymptotic results for the classical
heavy-traffic regime as described in Subsection~\ref{asym}
where the number of servers~$N$ is fixed and the relative load tends
to one in the limit.
The papers \cite{FS78,Reiman84,ZHW95} establish diffusion limits
for the JSQ policy in a sequence of systems with Markovian
characteristics as in our basic model set-up, but where in the $K$-th
system the arrival rate is $K \lambda + \hat\lambda \sqrt{K}$, while
the service rate of the $i$-th server is $K \mu_i + \hat\mu_i \sqrt{K}$,
$i = 1, \dots, N$, with $\lambda = \sum_{i = 1}^{N} \mu_i$,
inducing critical load as $K \to \infty$.
It is proved that for suitable initial conditions the queue lengths are
of the order O($\sqrt{K}$) over any finite time interval and exhibit
a state-space collapse property.

Atar {\em et al.}~\cite{AKM17} investigate a similar scenario,
and establish diffusion limits for three policies: the JSQ($d$) policy,
the redundancy-$d$ policy (where the redundant clones are abandoned
as soon as the first replica starts service), and a combined policy
called Replicate-to-Shortest-Queues (RSQ) where $d$ replicas are
dispatched to the $d$-shortest queues.

\paragraph{Non-degenerate slowdown.}

Asymptotic results for the so-called non-degenerate slow-down regime
described in Subsection~\ref{asym} where $N - \lambda(N) \to \gamma > 0$
as the number of servers~$N$ grows large, are scarce.
Gupta \& Walton~\cite{GW17} characterize the diffusion-scaled queue
length process under the JSQ policy in this asymptotic regime.
They further compare the diffusion limit for the JSQ policy with that
for a centralized queue as described above as well as several LBAs
such as the JIQ scheme and a refined version called Idle-One-First (I1F),
where a task is assigned to a server with exactly one task if no idle
server is available and to a randomly selected server otherwise.

It is proved that the diffusion limit for the JIQ scheme is no longer
asymptotically equivalent to that for the JSQ policy in this asymptotic
regime, and the JIQ scheme fails to achieve asymptotic optimality
in that respect, as opposed to the behavior in the large-capacity
and Halfin-Whitt regimes discussed in Subsection~\ref{ssec:jiq}.
In contrast, the I1F scheme does preserve the asymptotic equivalence
with the JSQ policy in terms of the diffusion-scaled queue length process,
and thus retains asymptotic optimality in that sense.

\paragraph{Sparse-feedback regime.}

As described in Section~\ref{ssec:jiq}, the JIQ scheme involves 
a communication overhead of at most one message per task, and yet
achieves optimal delay performance in the fluid and diffusion regimes.
However, even just one message per task may still be prohibitive,
especially when tasks do not involve big computational tasks,
but small data packets which require little processing.

Motivated by the above issues, \cite{BBL17b} proposes a novel class
of LBAs which also leverage memory at the dispatcher,
but allow the communication overhead to be seamlessly adapted
and reduced below that of the JIQ scheme.
Specifically, in the proposed schemes, the various servers provide
occasional queue status notifications to the dispatcher,
either in a synchronous or asynchronous fashion.
The dispatcher uses these reports to maintain queue estimates,
and forwards incoming tasks to the server with the lowest queue estimate.
The results in~\cite{BBL17b} demonstrate that the proposed schemes
markedly outperform JSQ($d$) policies with the same number of $d \geq 1$
messages per task and they can achieve a vanishing waiting time
in the limit when the update frequency exceeds $\lambda / (1 - \lambda)$.
In case servers only report zero queue lengths and suppress
updates for non-zero queues, the update frequency required for
a vanishing waiting time can in fact be lowered to just~$\lambda$,
matching the one message per task involved in the JIQ scheme.

\paragraph{Scaling of maximum queue length.}

So far we have focused on the asymptotic behavior of LBAs in terms
of the number of servers with a certain queue length, either on fluid
scale or diffusion scale, in various regimes as $N \to \infty$.
A related but different performance metric is the maximum queue length
$M(N)$ among all servers as $N \to \infty$.
Luczak \& McDiarmid~\cite{LM06} showed that for fixed $d \geq 2$ the
steady-state maximum queue length $M(N)$ under the JSQ($d$) policy is
given by $\log(\log(N)) / \log(d) + O(1)$ and is concentrated on at most
two adjacent values, whereas for purely random assignment ($d=1$),
it scales as $\log(N) / \log(1/\lambda)$ and does not concentrate
on a bounded range of values.
This is yet a further manifestation of the ``power-of-choice'' effect.

The maximum queue length $M(N)$ is the central performance metric
in balls-and-bins models where arriving items (balls) do not get
served and never depart but simply accumulate in bins,
and (stationary) queue lengths are not meaningful.
In fact, the very notion of randomized load balancing and power-of-$d$
strategies was introduced in a balls-and-bins setting in the seminal
paper by Azar {\em et al.}~\cite{ABKU94}.

%

\end{document}